\documentclass[runningheads]{llncs}
\usepackage{graphicx}
\usepackage{multirow}
\usepackage{color}
\usepackage{hyperref}
\usepackage{ulem}

\usepackage[symbol]{footmisc}
\renewcommand{\thefootnote}{\fnsymbol{footnote}}
\newcommand*\samethanks[1][\value{footnote}]{\footnotemark[#1]}

\begin{document}
\title{Text2Brain: Synthesis of Brain Activation Maps from Free-form Text Query}

\author{Anonymous\inst{1}}
\institute{Anonymous}

\author{Gia H. Ngo\inst{1}\thanks{indicates equal contribution} \and
Minh Nguyen\inst{2}\samethanks \and
Nancy F. Chen\inst{3}\thanks{indicates equal contribution} \and
Mert R. Sabuncu\inst{1,4}\samethanks}
\authorrunning{Ngo \& Nguyen et al.}

\institute{School of Electrical \& Computer Engineering, Cornell University, USA \and
Computer Science Department, University of California, Davis, USA \and
Institute of Infocomm Research (I2R), A*STAR, Singapore\and
Radiology, Weill Cornell Medicine, USA
}

\maketitle              %
\renewcommand*{\thefootnote}{\arabic{footnote}}
\begin{abstract}
Most neuroimaging experiments are under-powered, limited by the number of subjects and cognitive processes that an individual study can investigate.
Nonetheless, over decades of research, neuroscience has accumulated an extensive wealth of results.
It remains a challenge to digest this growing knowledge base and obtain new insights since existing meta-analytic tools are limited to keyword queries.
In this work, we propose Text2Brain, a neural network approach for coordinate-based meta-analysis of neuroimaging studies to synthesize brain activation maps from open-ended text queries.
Combining a transformer-based text encoder and a 3D image generator, Text2Brain was trained on variable-length text snippets and their corresponding activation maps sampled from 13,000 published neuroimaging studies.
We demonstrate that Text2Brain can synthesize anatomically-plausible neural activation patterns from free-form textual descriptions of cognitive concepts.
Text2Brain is available at \url{https://braininterpreter.com} as a web-based tool for retrieving established priors and generating new hypotheses for neuroscience research.

\keywords{coordinate-based meta-analysis  \and transformers \and information retrieval \and image generation.}
\end{abstract}
\section{Introduction}
\label{sec:intro}
Decades of neuroimaging research have yielded an impressive repertoire of findings and greatly enriched our understanding of the cognitive processes governing the mind.
However, individual brain imaging experiments are often under-powered~\cite{carp2012secret,button2013power}, constrained by the number of subjects and psychological processes that each experiment can probe~\cite{church2010task}.
To synthesize reliable trends across such experiments, researchers often perform meta-analysis on the coordinates of the most significant effect (such as 3D location of peak brain activation in response to a task).
Most meta-analyses require the expert selection of relevant experiments (e.g. \cite{costafreda2008predictors,minzenberg2009meta,shackman2011integration}).
One key challenge with conducting meta-analysis on neuroimaging experiments is the consolidation of synonymous terms.
As neuroscientific research constantly evolves, different denominations might be used in different contexts or invented to refine existing ideas.
For instance, ``self-generated thought'', one of the most highly studied functional domains of the human brain~\cite{smallwood2013distinguishing}, can be referred to by different terms such as ``task-unrelated thought''~\cite{andrews2014default}.

Manual selection of experiments for meta-analysis can be replaced by automated keyword search through data automatically scraped from the neuroimaging literature~\cite{yarkoni2011large,dockes2020neuroquery,rubin2017decoding}.
For example, Neurosynth~\cite{yarkoni2011large} and more recently Neuroquery~\cite{dockes2020neuroquery} both use automated keyword search to retrieve relevant studies to synthesize brain activation maps from text queries.
However, Neurosynth and Neuroquery only allow for rigid queries formed out of predefined keywords and rely on superficial lexical similarity via co-occurrences of keywords for inference of longer or rarer queries.
We propose an alternative approach named Text2Brain, which permits more flexible free-form text queries.
Text2Brain also characterizes more fine-grained and implicit semantic similarity via vector representations from neural modeling in order to retrieve more relevant studies.
Moreover, existing approaches estimate voxel-wise activations using either univariate statistical testing or regularized linear regression.
In contrast, Text2Brain generates whole-brain activation maps using a 3D convolutional neural network (CNN) for more accurate construction of both coarse and fine details.

We compare Text2Brain's predictions with those from established baselines where we used article titles as free-form queries. %
Furthermore, we assess model predictions on independent test datasets, including reliable task contrasts and meta-analytic activation maps of well-studied cognitive domains predicted from their descriptions.
Our analysis shows that Text2Brain generates activation maps that better match the target images than the baselines do.
Given its flexibility in taking input queries, Text2Brain can be used as an educational aid as well as a tool for synthesizing prior maps for future research.

\section{Materials and Methods}
\subsection{Overview}
\begin{figure}
    \centering
    \includegraphics[width=0.9\linewidth]{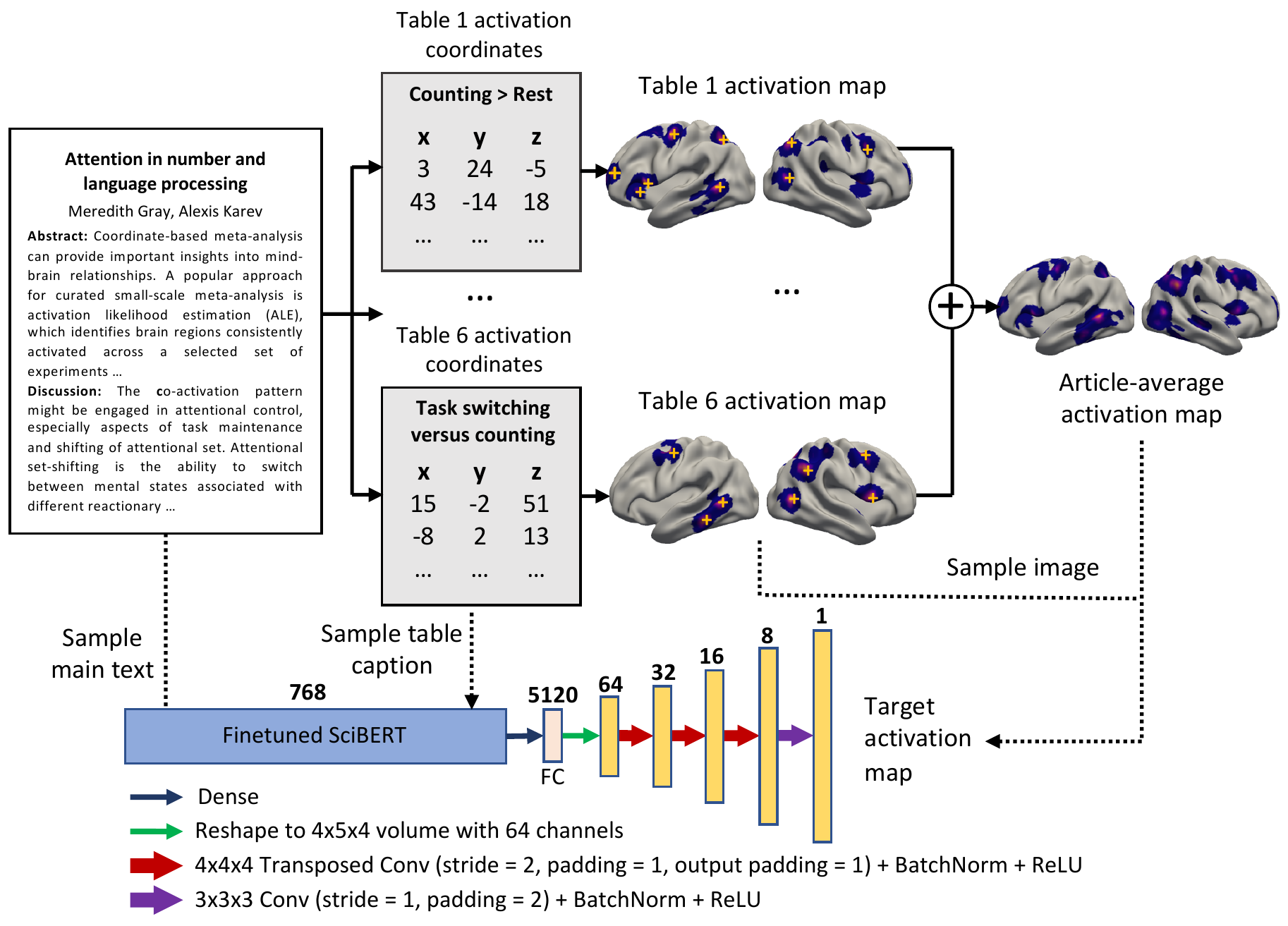}
    \caption{Overview of data preprocessing, the Text2Brain model, and training procedure. All activation maps are 3D volumes, but projected to the surface for visualization.}
    \label{fig:model}
\end{figure}

Figure \ref{fig:model} shows the  overview of our approach.
For each research article, full text and activation coordinates are extracted to create training samples (section \ref{subsec:data}).
Text2Brain model consists of a transformer-based text encoder and a 3D CNN~(section \ref{subsec:model}).
The transformer uses attention to encode the input text into vector representation~\cite{vaswani2017attention,devlin2018bert}.
Thus, over many text-brain activation map pairs, the model automatically learns the association between activation at a spatial location with the most relevant words in the input text.
Unlike classical keyword search that mainly exploits co-occurrence of keywords regardless of context, a transformer refines the vector representation depending on the specific phrasing of the text inputs (i.e. context)~\cite{tenney2019you}.
This allows Text2Brain to map synonymous text to a similar activation map.
Instead of explicitly searching through articles, Text2Brain stores the articles' content in its parameters~\cite{petroni2019language} and outputs a relevant vector representation when presented with an input query.
Thus, we use an augmented data sampling strategy to encourage the model to construct and store rich many-to-one mappings between textual description and activation maps (section \ref {subsec:training}).

\subsection{Data Preprocessing}
\label{subsec:data}
Coordinates of peak activation were scraped from tables of results reported in more than 13,000 neuroimaging articles and previously released in \cite{dockes2020neuroquery}.
Each table has the corresponding article's PubMed ID, the table's ID as originally numbered in the article, and coordinates of peak activation converted to MNI152 coordinate system \cite{lancaster2007bias}.
Following the preprocessing procedure of \cite{dockes2020neuroquery}, a Gaussian sphere with full width at half maximum (FWHM) of 9mm is placed at each of the coordinates of peak activation.
Thus, an activation map for each table in an article is generated from the set of activation foci associated with the table.
An article-average activation map is also generated by averaging the activation maps of all the tables in the article.
The articles' full text are scraped using their PubMedID via NCBI API~\footnote{\url{https://www.ncbi.nlm.nih.gov/books/NBK25501/}} and Elsevier E-utilities API~\footnote{\url{https://dev.elsevier.com/}}.

\subsection{Model}
\label{subsec:model}

Figure \ref{fig:model} shows the basic schematic of Text2Brain, which consists of a text encoder based on SciBERT~\cite{beltagy2019scibert} and a 3D CNN as the image decoder.
Output embedding of the text encoder is of dimension $768$ and projected via a fully-connected layer, then reshaped to a 3D volume of dimension $4\times5\times4$ and $64$ channels at each voxel.
The image decoder consists of 3 transposed 3D convolutional layers with 32, 16, 8 channels respectively.
The model was trained using mean-squared error for 2000 epochs, batch size of 24  with Adam \cite{loshchilov2018decoupled}.
The learning rate for the text encoder and image decoder are $10^{-5}$ and $3\times{10^{-2}}$, respectively.
The model's source code is available at \url{https://github.com/sabunculab/text2brain}.

\subsection{Training}
\label{subsec:training}
During training, an activation map is sampled with equal probability from the set of table-specific activation maps and the article-average map.
For each table-specific activation map, the first sentence of the table caption (as our data exploration suggested this to be the most useful description) is also extracted as the image's corresponding text.
For each article-average activation map, one of four types of text is sampled with equal probability as the approximate description of the activation pattern, namely (1) the article's title (2) one of the article's keywords (2) abstract (3) a randomly chosen subset of sentences from the discussion section.
This augmented sampling strategy encourages Text2Brain to generalize over input texts of different lengths.
Furthermore, sampling multiple text snippets for an activation pattern encourages the model to automatically infer keywords present across queries and implicitly learn the association between different but synonymous words with an activation map.
Supplemental Figure~\ref{fig:supp_ablation} shows an ablation study on the sampling strategy.

\section{Experimental Setup}
\subsection{Predict activation maps from article titles}

From the dataset of 13000 articles, 1000 articles are randomly sampled as the test set such that the keywords (defined by the articles' authors) are not included in the training and validation articles.
Of the remaining articles, 1000 are randomly held out as a validation set for parameters tuning.
For each article, the article-average activation map is predicted from its title using Text2Brain and the two baselines of Neurosynth and Neuroquery.

\subsection{Predict activation maps from contrast descriptions}
The Human Connectome Project (HCP) offers neuroimaging data from over 1200 subjects, including task fMRI (tfMRI) of 86 task contrasts from 7 domains~\cite{barch2013function}.
While detailed descriptions of task contrasts are provided by HCP, we instead use the more concise contrast descriptions provided by the Individual Brain Charting (IBC) project~\cite{pinho2020individual}, which includes fMRI data from 12 subjects and 180 task contrasts, 43 of which are also studied in the HCP.
The reason for using the IBC contrast descriptions is because they are more succinct and thus more favorable to the baselines.
The target (ground-truth) activation maps are the group-average contrast maps provided by the HCP, as the large number of subjects provides more reliable estimates of the contrast maps.
In our analyses, we use the agreement between the IBC and HCP maps as a measure of reliability.
Note that despite using similar experimental protocols, there are subtle differences between the IBC and HCP experiments.
For example, while the original HCP language task was conducted in English, the corresponding language task in the IBC project was conducted in French.
Overall, Text2Brain and the two baselines were evaluated on the 43 HCP task contrasts.

\subsection{Baselines}
The first baseline, Neurosynth~\cite{yarkoni2011large}, collected all peak activation coordinates across neuroimaging articles that mention a given keyword and performed a statistical test at every voxel to determine a significant association. %
For longer query, we performed statistical test using activation coordinates reported in all articles that contain at least one of the keywords in the input text.

The second baseline, Neuroquery~\cite{dockes2020neuroquery}, builds upon Neurosynth by extending the vocabulary of keywords via manual selection from other sources. The keyword encoding is obtained after performing non-negative matrix factorization of the articles' abstract (as a bag of keywords) represented with {term frequency - inverse document frequency (TF-IDF) features~\cite{salton1988term}. A ridge regression model was trained to learn the mapping from the text encoding to the activation at each voxel. The inference of a keyword is smoothed by a weighed summation with most related keywords (in the TF-IDF space). For longer queries, the predicted activation is the average of maps predicted from all keywords in the input.

\subsection{Evaluation Metrics}
To measure the similarity of predicted and target activation maps at different levels of detail, we compute Dice scores \cite{dice1945measures} at various thresholds. %
This evaluation procedure is similar to that used in \cite{dockes2020neuroquery} for a thresholded target map, but we apply the same thresholding to both the target and predicted map.
For example, at a lower threshold (e.g., considering the 5\% most activated voxels), the Dice score measures the correspondence of the fine-grained details between the target and predicted activation maps.
At higher thresholds (e.g. 25\% most activated voxels), this metric captures gross agreement of activation clusters.
We also compute an approximated integration of Dice scores across all thresholds (from 5\% up to 30\%), i.e. the area under the Dice curve (AUC), as a summary measure.
Supplemental Figure \ref{fig:supp_dice} shows the Dice curve for an example pair of target-predicted activation maps.
We only consider up to 30\% to be fair to the baselines, as the portion of activated voxels predicted by Neuroquery only extends up to 30\% of the gray matter mask.

\section{Results}
\label{sec:results}
\subsection{Validate activation maps predicted from article title}
\begin{figure}
    \centering
    \includegraphics[width=0.8\linewidth]{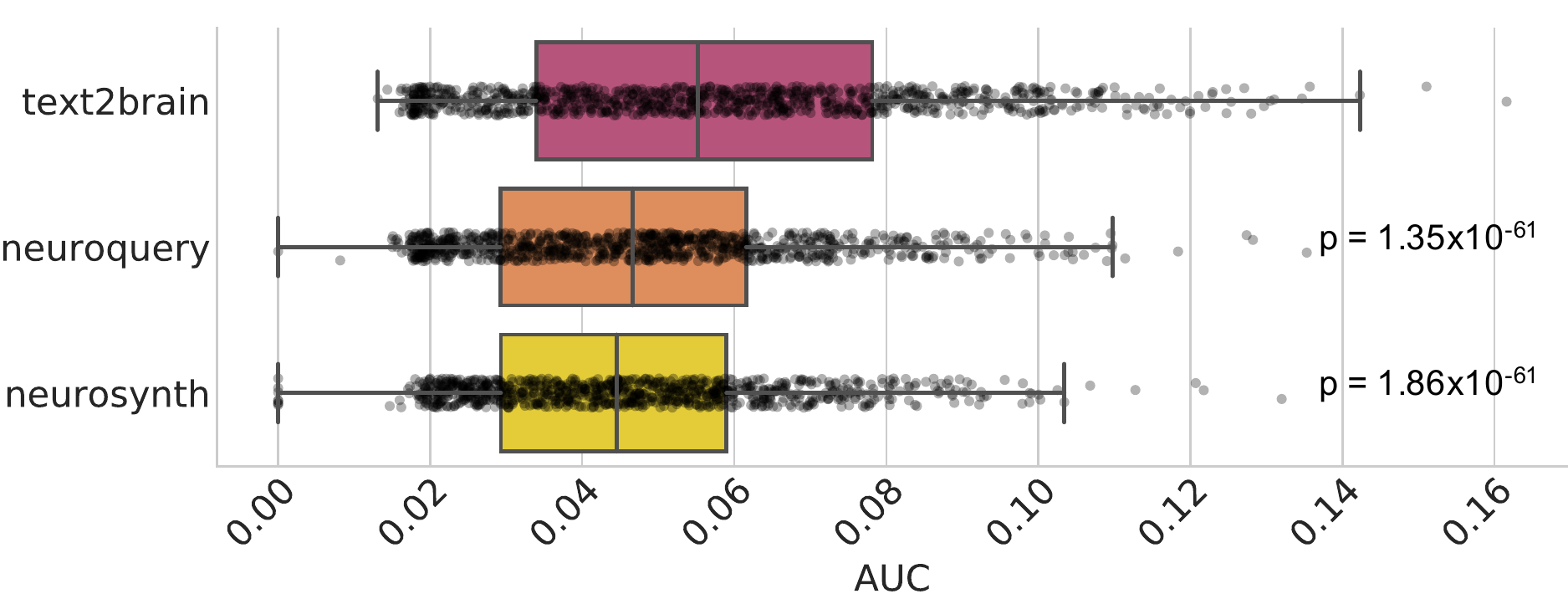}
    \caption{Evaluation of article-average activation maps predicted from their titles measured in area under the Dice curve (AUC) score. The p-values are computed from paired-sample t-tests between Text2Brain and each of the 2 baselines.}
    \label{fig:article_AUC}
\end{figure}
Figure \ref{fig:article_AUC} compares the quality of activation maps predicted from the titles of 1000 articles. %
Text2Brain model (mean AUC = $0.0576$) outperforms Neuroquery (mean AUC = $0.0478$) and Neurosynth (mean AUC = $0.0464$).
Paired-sample t-tests show that this performance gap is statistically very significant. 
The p-value for the comparison between Neuroquery and Neurosynth is $p=0.015$.
While Text2Brain can make a prediction for all samples, Neurosynth and Neuroquery fail to make prediction for some article titles, resulting in zero AUCs values.

\subsection{Prediction of task contrast maps from description}
\begin{figure}
    \centering
    \includegraphics[width=\linewidth]{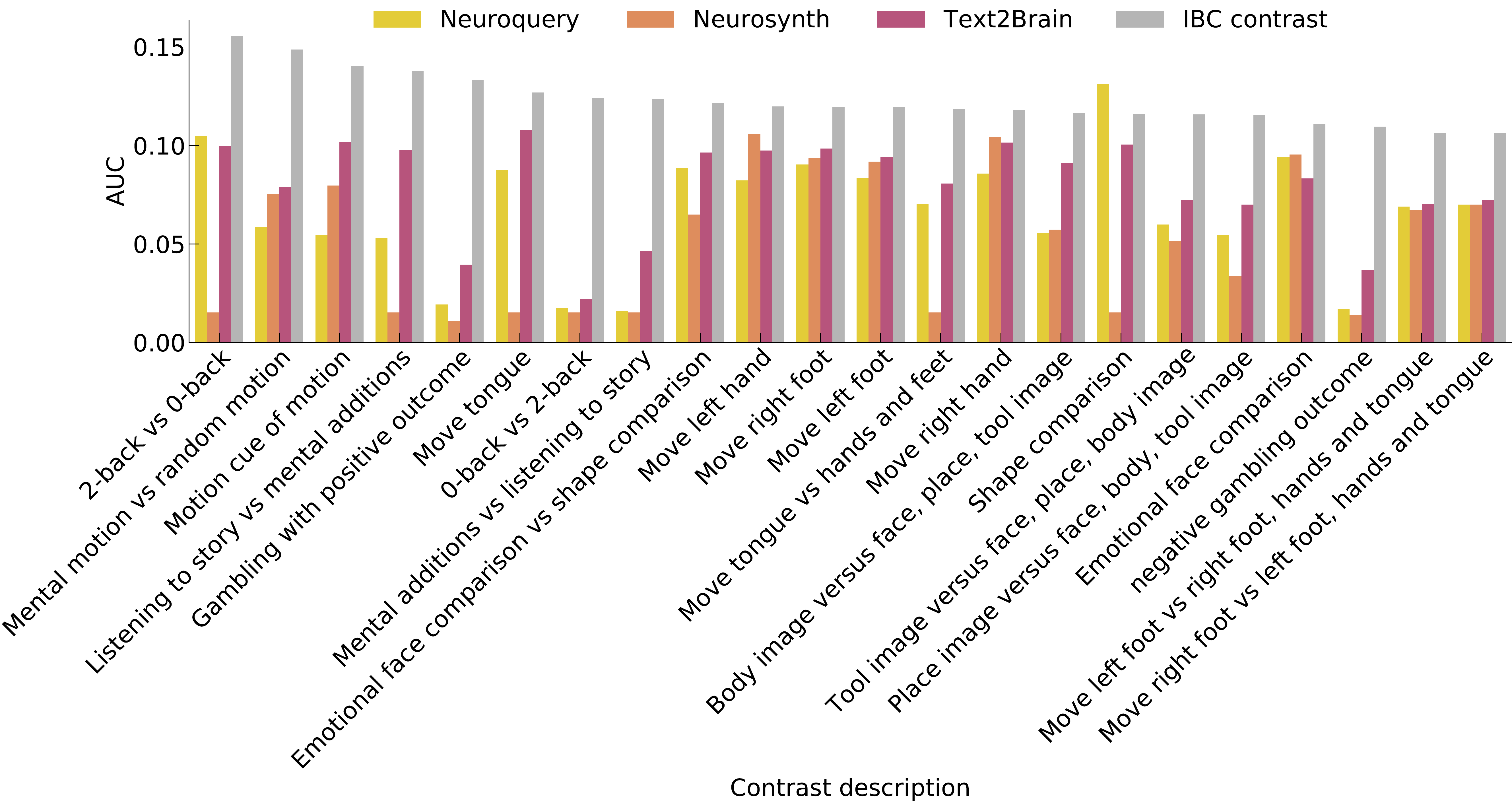}
    \caption{AUC of predicted HCP task activation maps from contrasts' description.}
    \label{fig:hcp_auc}
\end{figure}
Figure \ref{fig:hcp_auc} shows the AUC scores for the prediction of the three models and the IBC average contrasts, against the HCP target maps.
The 22 contrasts with the HCP-IBC's AUC score above the average, considered to be the reliable contrasts, are shown.
Across all 43 HCP contrasts, Text2Brain (mean AUC = $0.082$) performs better than the baselines, i.e. Neuroquery (mean AUC = $0.0755$, $p = 0.08$), Neurosynth (mean AUC = $0.047$, $p = 1.5\times10^{-5}$), where $p$-values are computed from the paired t-test between Text2Brain's and the baselines' prediction.
As reference, IBC contrasts yield mean AUC = $0.094$ ($p = 0.077$).
\begin{figure}
    \centering
    \includegraphics[width=\linewidth]{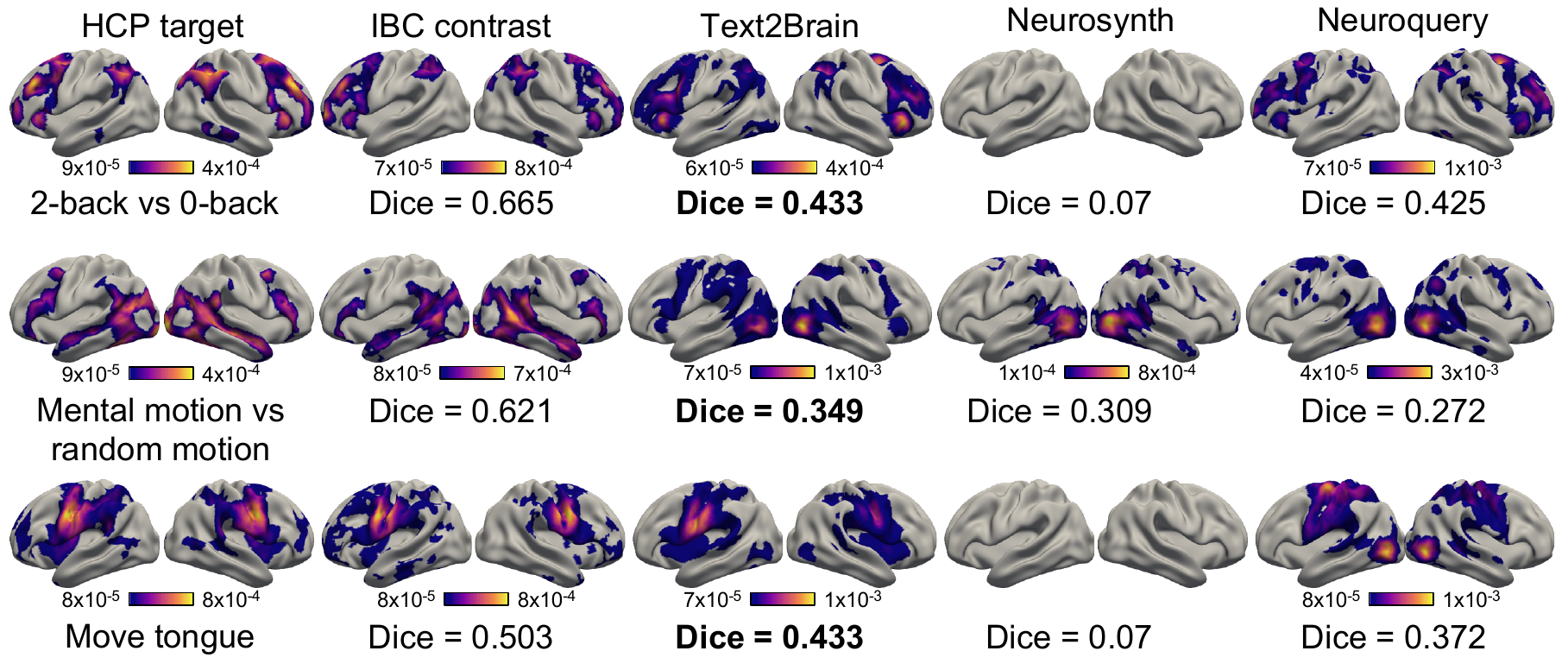}
    \caption{Task activation maps predicted from contrasts' description. The Dice scores are computed between the binarized map of 20\% most activated voxels in the predicted and target brain maps.}
    \label{fig:hcp_example}
\end{figure}

Figure \ref{fig:hcp_example} shows the prediction from three most reliable task contrasts (having the highest HCP-IBC AUC), thresholded at the top 20\% most activated voxels.
The three contrasts correspond to different HCP task groups, namely ``WORKING MEMORY'', ``SOCIAL'', and ``MOTOR''.
Text2Brain's prediction improves over the baselines for the three contrasts.
Neurosynth was not able to generate activation maps for two of the contrast descriptions (``2-back vs 0-back'' and ``Move tongue'').
On the other hand, for the ``Move tongue'' contrast, Neuroquery predicts activation in the primary cortex, but the peak is in the wrong location, shifted more toward the hand region of the homunculus. Additionally, there is a false positive prediction in the occipital cortex.

Finally, we are interested in examining the prediction for ``Self-generated thought'', which is one of the most commonly studied functional domains, due to its engagement in a wide range of cognitive processes that do not require external stimuli~\cite{andrews2014default},
and is associated with the default network~\cite{buckner2008brain}.
The ground-truth map for self-generated thought, taken from~\cite{ngo2019beyond}, is estimated using activation likelihood estimation (ALE)~\cite{turkeltaub2002meta,laird2005ale,eickhoff2009coordinate}, a well established tool for coordinate-based meta-analysis, applied on 1812 activation foci across 167 imaging studies over 7 tasks based on strict selection criteria~\cite{spreng2009common,mar2011neural,sevinc2014contextual}.
\begin{figure}
    \centering
    \includegraphics[width=\linewidth]{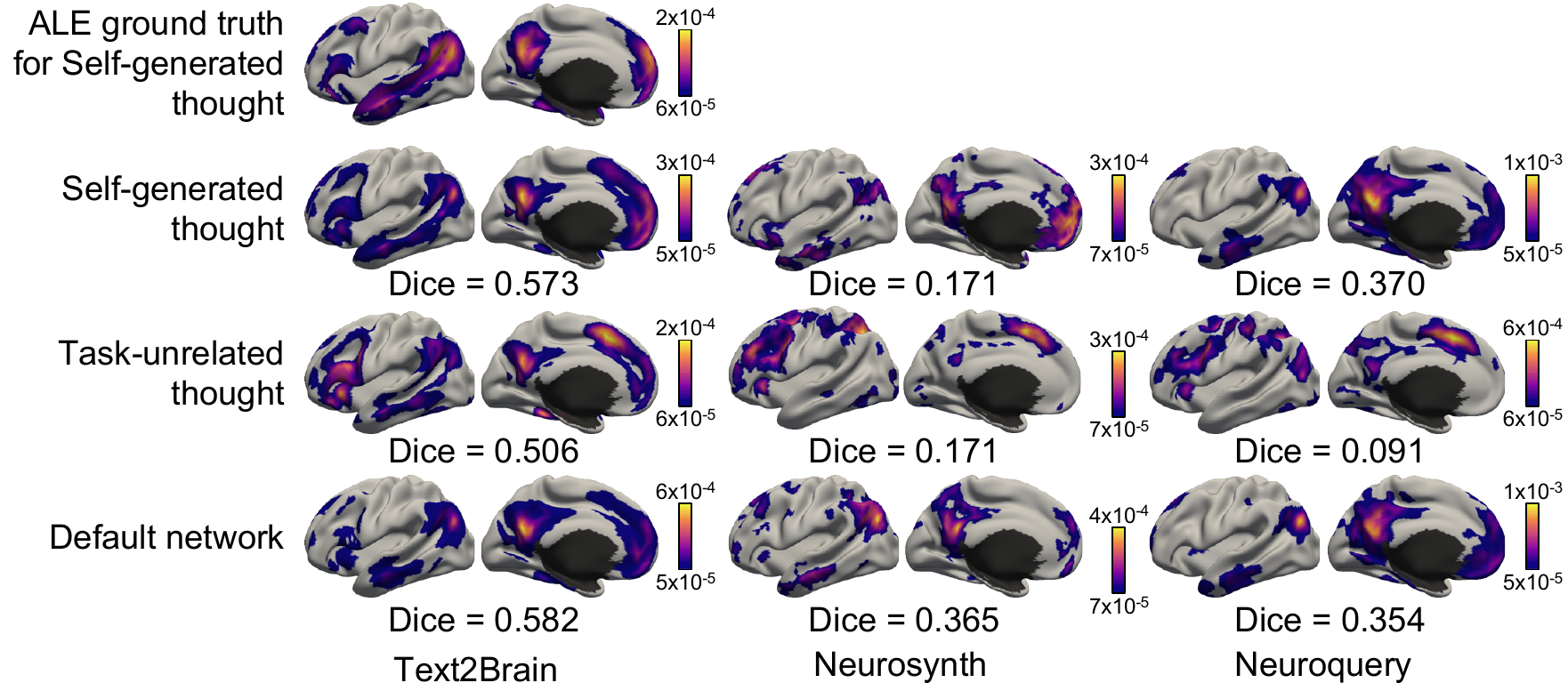}
    \caption{Prediction of self-generated thought activation map using synonymous queries}
    \label{fig:self_generated_thought}
\end{figure}
Figure \ref{fig:self_generated_thought} shows the prediction of self-generated thought activation map using three different query terms, thresholded at the top 20\% most activated voxels.
For the ``self-generated thought'' and ``default network'' queries, all approaches generate activation maps that are consistent with the ground-truth, which includes the precuneus, the medial prefrontal cortex, the temporo-parietal junction, and the temporal pole.
Text2Brain's prediction best matches the ground-truth activation map compared to the baselines.
Text2Brain can also replicate a similar activation pattern from the query ``task-unrelated thought'', evident by only a slight drop in the Dice score.
However, Neuroquery and Neurosynth both produce activation maps that deviate from the typical default network's regions with increased activation in the prefrontal cortex, also evident by a large drop in the Dice scores.

\section{Conclusion}
In this work, we present a model named Text2Brain for generating activation maps from free-form text query.
By finetuning a high-capacity SciBERT-based text encoder to predict coordinate-based meta-analytic maps, Text2Brain captures the rich relationship in the language representational space, allowing the model to generalize its prediction for synonymous queries.
This is evident in the better performance of Text2Brain in predicting the self-generated thought activation map using different descriptions of the functional domain.
Text2Brain's capability to implicitly learn relationships between terms and images will help the model stays relevant and useful even as neuroimaging literature continues to evolve with new information and rephrasing of existing concepts.
We also show that Text2Brain accurately predicts most of the task contrasts included in the HCP dataset based on their description, validating its capability to make prediction for longer, arbitrary queries.
Text2Brain also avoids the failure cases suffered by Neurosynth and Neuroquery in which they cannot predict if the input words are not defined in their vocabularies, even though the queries are relevant to neuroscience research such as the title of an article or a contrast description.
In the future, we will work on the interpretability of the approach, such as to attribute regions of activation in the generated map to specific word in the input query, as well as to efficiently match activation maps and research text most relevant to the synthesized images.

\section*{Acknowledgement}
This work was supported by NIH grants R01LM012719, R01AG053949, the NSF NeuroNex grant 1707312, the NSF CAREER 1748377 grant and Jacobs Scholar Fellowship.

\bibliographystyle{unsrt}
\bibliography{references}

\appendix
\newpage
\pagenumbering{arabic}
\setcounter{page}{1}
\renewcommand{\thefigure}{S\arabic{figure}}
\setcounter{figure}{0}
\begin{center}
\large\textbf{Supplementary Materials for ``Text2Brain: Synthesis of Brain Activation Maps from Free-form Text Query''}
\end{center}
\section{Evaluation Metrics}
Dice score~\cite{dice1945measures} is used to measure the extent of overlap between a predicted activation map and the target activation map at a given threshold.
At a given threshold of $x\%$, Dice score is computed as:
\begin{equation}
    Dice(x) = \frac{2|Prediction(x) \cap Target(x)|}{|Prediction(x)|+|Target(x)|},
    \label{eq:dice}
\end{equation}
where $|Prediction(x)|$ denotes the number of top $x\%$ most activated voxels in the predicted activation map, $|Target(x)|$ denotes the number of top $x\%$ most activated voxels in the target map, and $|Prediction(x)\cap Target(x)|$ denotes the number of voxels that overlap between the predicted and target map at the given threshold.
\begin{figure}
\includegraphics[width=\textwidth]{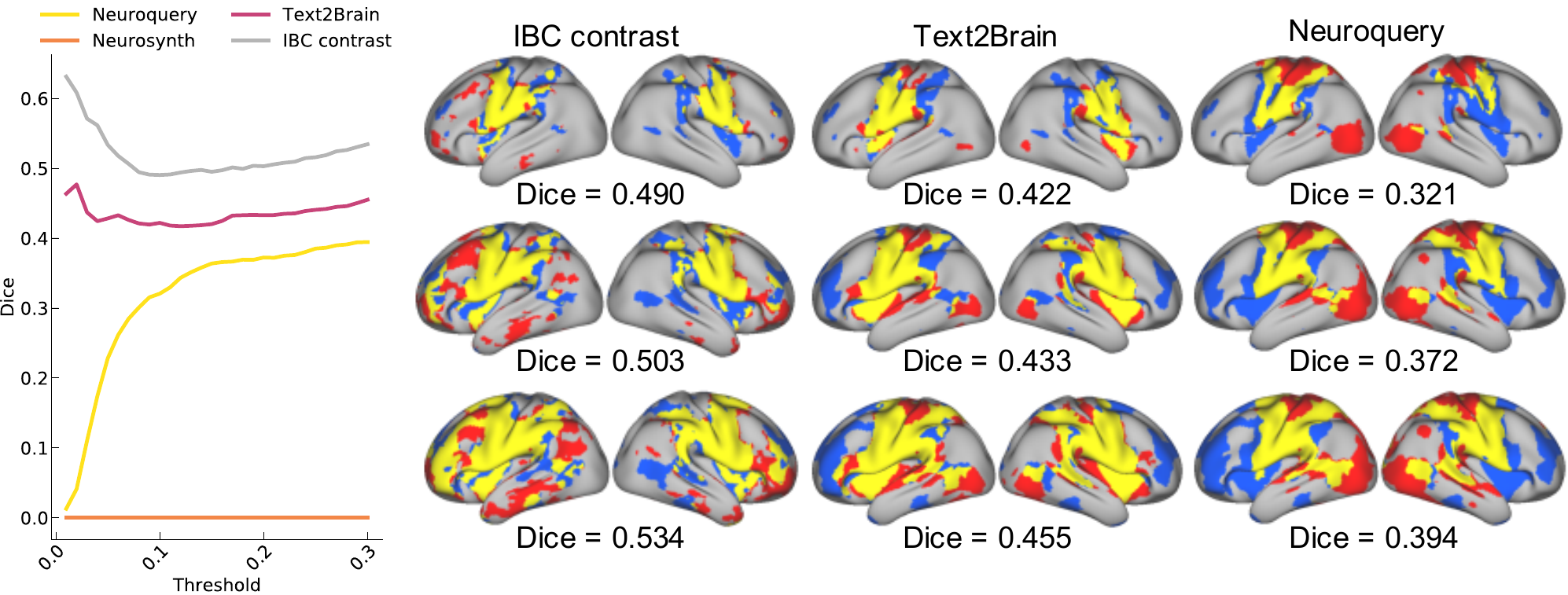}
\caption{Example Dice scores evaluated on the ``Move tongue'' contrast (in Fig. \ref{fig:hcp_example}.
The graph on the left shows the Dice scores computed between the target HCP activation map and Text2Brain's, Neurosynth's, Neuroquery's prediction, and the IBC contrast across thresholds ranging from 5\% to 30\%.
Note that Neurosynth's Dice scores are all zeros as it fails to make an inference for the input text.
The area under the Dice curve (AUC) was computed as the summary metrics of accuracy across all thresholds (e.g. Fig~\ref{fig:hcp_auc}).
The brain maps on the right are visualization of the extent of overlaps between predicted and target maps at 10\%, 20\% and 30\% threshold of most activated voxels. Blue indicates activation in the target contrast, red is the predicted activation and yellow is the overlap.
}\label{fig:supp_dice}
\end{figure}
\clearpage
\section{Ablation study of sampling strategy}
\label{sec:ablation}
\begin{table}[h!]
\centering
\begin{tabular}{ | c | c | } 
\hline
 Text samples                                             & Mean AUC \\ \hline
 Title + Table caption                                    & 0.0648    \\
 Title + Abstract + Table caption                         & 0.0616    \\
 Discussion + Abstract                                    & 0.0631    \\
 Discussion + Abstract + Keywords                         & 0.0651    \\
 \textbf{Title + Abstract + Keywords + Discussion + Table caption} & 0.0663     \\
 \hline
\end{tabular}
\end{table}
\vspace{-1cm}
\begin{figure}
 \includegraphics[width=\textwidth]{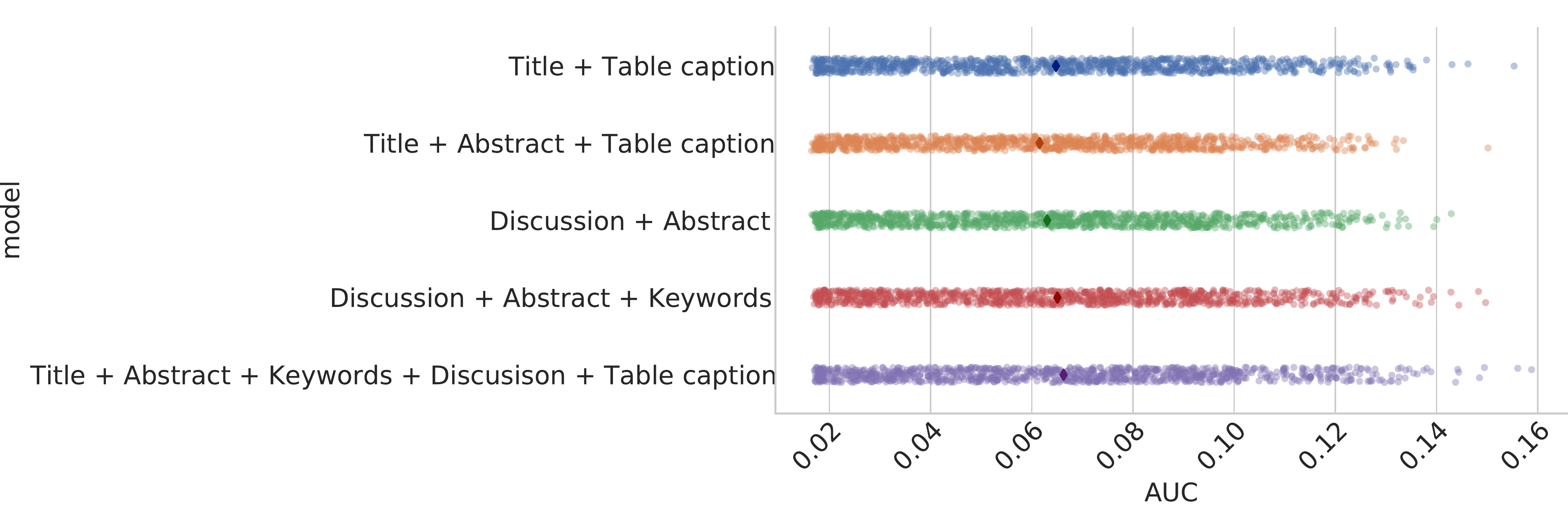}
 \caption{Performance of different sampling strategies in predicting article-average activation maps from the articles' titles in the validation set.
 All sampling strategies used the same model described in~\ref{subsec:model}.
 The model parameters used for evaluation were chosen at the epoch with the best performance on the validation set.}
 \label{fig:supp_ablation}
\end{figure}

\end{document}